**Scaling of Geographic Space from the Perspective of City and Field Blocks and Using Volunteered Geographic Information**

Bin Jiang and Xintao Liu

Department of Technology and Built Environment, Division of Geomatics
University of Gävle, SE-801 76 Gävle, Sweden
Email: bin.jiang@hig.se, xintao.liu@hig.se

**Abstract**
Scaling of geographic space refers to the fact that for a large geographic area its small constituents or units are much more common than the large ones. This paper develops a novel perspective to the scaling of geographic space using large street networks involving both cities and countryside. Given a street network of an entire country, we decompose the street network into individual blocks, each of which forms a minimum ring or cycle such as city blocks and field blocks. The block sizes demonstrate the scaling property, i.e., far more small blocks than large ones. Interestingly, we find that the mean of all the block sizes can easily separate between small and large blocks- a high percentage (e.g., 90%) of smaller ones and a low percentage (e.g., 10%) of larger ones. Based on this regularity, termed as the head/tail division rule, we propose an approach to delineating city boundaries by grouping the smaller blocks. The extracted city sizes for the three largest European countries (France, Germany and UK) exhibit power law distributions. We further define the concept of border number as a topological distance of a block far from the outmost border to map the center(s) of the country and the city. We draw an analogy between a country and a city (or geographic space in general) with a complex organism like the human body or the human brain to further elaborate on the power of this block perspective in reflecting the structure or patterns of geographic space.

**1. Introduction**
Scaling of geographic space refers to the fact that for a large geographic area the small constituents or units are much more common than the large ones (Jiang 2010). For example, there are far more short streets than long ones (Kalapala et al. 2006, Jiang 2007); far more small city blocks than large ones (Lämmer et al. 2006); far more small cities than large ones, a phenomenon referred to as Zipf's law (1949); far more short axial lines than long ones (Jiang and Liu 2010). This notion of far more small things than large ones is a de facto heavy tailed distribution that includes power law, exponential, lognormal, stretched exponential, and power law with a cutoff (Clauset et al. 2009). The heavy tailed distributions have been well studied to characterize many natural and societal phenomena, and have received a revival of interest in the Internet age; interested readers can refer to Clauset et al. (2009) and references therein for more details. The heavy tailed distributions differ fundamentally from a normal distribution. For things that follow a normal distribution, the sizes of the things would not vary much but be centered around a typical value – a mean or an average. In this respect, the standard deviation is used to measure the overall variation between the individual values and the mean. The scaling of geographic space is closely related to, yet fundamentally different from, the conventional concept of spatial heterogeneity, which is usually characterized by a normal distribution (Anselin 2006, Goodchild 2004).

This paper is motivated by the belief that geographic space essentially exhibits a heavy tailed distribution rather than a normal distribution. We attempt to investigate the scaling of geographic space from the perspective of city and field blocks using street networks involving both cities and countryside of the three largest European countries: France, Germany and UK. We take the three street networks and decompose them into individual blocks, each of which forms a minimum ring or cycle such as city blocks and field blocks. The sizes of the blocks in each country exhibit a lognormal distribution, i.e., far more small blocks than large ones. We further find that the mean of the blocks make a clear-cut distinction between small and large blocks, and that the blocks with a size below the mean tend to belong to city blocks, while those above the mean belong to field blocks. In contrast to city blocks, field blocks are surrounded by a minimum ring of road segments in countryside. Based on this interesting finding and using the concept of spatial autocorrelation, we develop an approach to defining and identifying cities or city boundaries for countries. We define the border number as the topological distance of blocks far from the outmost borders extracted from the street networks; refer to Section 2 for more details. We map the border number for individual blocks to identify true center(s) of the countries.

This paper is further motivated by another intriguing issue, i.e., how to delineate city boundaries. Delineating city boundaries objectively is essential for many urban studies and urban administrations (Rozenfeld et al. 2008). Researchers and practitioners alike usually rely on the boundaries provided by census or statistical bureaus. These imposed boundaries are considered to be subjective or even arbitrary. Advanced geospatial technologies like GIS and remote sensing provide updated data sources for extracting urban boundaries. However, to the best of our knowledge, existing solutions are established on a raster format. For example, detect urban areas from satellite imagery in particular using nightlight imagery (Sutton 2003) and create city boundaries from density surface of street junctions based on kernel density estimation (Borruso 2003, Thurstain-Goodwin and Unwin 2000). Jiang and Jia (2011) adopted an approach that clusters street nodes (including intersections and ends) first and then imposes a grid to delineate city boundaries, thus being raster-based in essence. The raster-based solutions, due to the resolution choice, inevitably suffer from the modified areal unit problem – the sizes of map units affect the results of statistical hypothesis tests (Openshaw 1984). Following scaling analysis of city and field block sizes, this paper suggests a vector-based approach to delineating city boundaries. This approach is considered to be a by-product of the scaling analysis, a nice application of the head/tail division rule derived (c.f., Section 3 for more details).

The contribution of this paper is three-fold: (1) we find that the mean of block sizes can divide all blocks into city blocks and field blocks; (2) based on this finding, we develop a novel approach to delineating city boundaries from large street networks; and (3) we define border number from a topological perspective to map the centers of a large geographic space. Besides, we provide a set of procedures or algorithms to extract city and field blocks from large street networks using data-intensive geospatial computing.

The remainder of this paper is structured as follows. Section 2 introduces in a detailed manner data and data processing to extract individual city and field blocks from the large street networks of the three European countries. Section 3 illustrates the lognormal distribution of block sizes, based on which we derive the head/tail division rule to characterize the inbuilt imbalance between a minority of blocks in the head and a majority of those in the tail. Based on the head/tail division rule, in Section 4, we develop an approach to delineating urban boundaries that are naturally defined, termed by natural cities. The sizes of the natural cities follow a power law-like distribution, $P(x) \sim x^{-\alpha}$, one of the heavy tailed distributions. Section 5 adopts the concept of border number to map the center(s) of countries or cities. Before drawing a conclusion to this paper in Section 7, we elaborate on the implications of the study in Section 6, in particular on how a country, a city or geographic space in general can be compared to a complex organism in terms of the imbalanced structure and the self-organization processes.

## 2. Data and data processing
Data and data processing constitutes a very important part of the research. The data is taken from OpenStreetMap (OSM, www.openstreetmap.org), a wiki-like collaboration or grass roots movement which aims to create a free editable map of the world using free sources like GPS traces and digital imagery (Haklay and Weber 2008, Bennett 2010). The OSM data is one of many successful examples of volunteered geographic information contributed by individuals and supported by the Web 2.0 technologies (Goodchild 2007). The quantity and quality of OSM data can be compared to that of data collected by national mapping agencies or commercial companies.

### 2.1 Data pre-processing for building up topological relationships
We adopt the street networks of three largest European countries (France, Germany, and UK) for the computation and experiments. Before the extraction of individual blocks for scaling analysis, we need to build up topological relationships. This is because the original OSM data are without topology, much like digitizing lines without generating coverage – a topology-based vector data format. Through this pre-processing, all line segments will be assigned a direction and become arcs that meet at nodes and have left and right polygons. The original OSM data are too large to load into any existing geographic information systems (GIS) package for the pre-processing. To solve this problem we parceled them into several pieces, processed them separately and merged them again afterwards. To have some idea of the data sizes, the numbers of arcs for the individual countries are listed in Table 1. As one can see, there are several million arcs for each of the three networks, up to 8 million arcs for the Germany street network. It took several hours for a 64 bits machine (4 cores CPU, 48 GB memory and 1TB hard disk) to finish the pre-processing of creating the topological relationships for each of the three networks.

Table 1: The number of arcs, blocks and the maximum border number for the three street networks

|  | France | Germany | UK |
| --- | --- | --- | --- |
| Arcs | 2,323,980 | 8,176,518 | 2,970,534 |
| Blocks | 569,739 | 2,095,388 | 586,809 |
| MaxBorder# | 79 | 125 | 73 |

**2.2 Extraction of city and field blocks**

Based on the pre-processing, we compute the arc-based networks to extract individual blocks in order to investigate some scaling properties. To introduce the computation, we adopt a fictive street network shown in Figure 1, which includes forty blocks and several dangling arcs that do not constitute any part of the blocks. To extract the individual blocks, we first need to set a minimum bounding box for the network in order to select an outmost arc to start traversal processes. There are two kinds of traversal processes: left traversal process and right traversal process. The left traversal process means that when comes to a node with two or more arcs, it always chooses the most left arc. On the other hand, it always chooses the most right arc for the right traversal process. Once the traversal process (starting from the outmost arc) is over, it ends up with one cycle: either a minimum cycle (which is a block) or a maximum cycle which is the outmost border. If the maximum cycle is not generated, then the program chooses a reverse direction for the traversal process until the maximum cycle is detected, and the corresponding arcs are marked with the traversal direction (left or right).

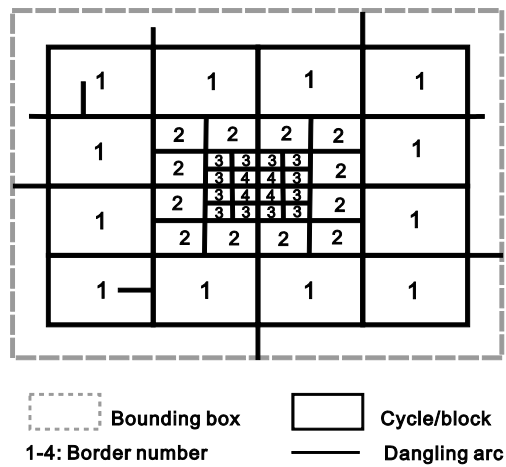

Figure 1: Illustration of the minimum cycles (or blocks) and the maximum cycle

The next step is to choose an arc on the border, and begin the traversal process along the opposite direction as previously marked, until all arcs on the border are processed. This way all the blocks on the border are detected and are assigned to border number 1. This process goes recursively for the blocks that are adjacent to the blocks with border number 1. We will get all the blocks with border number 2. The above process continues until all the blocks are exhausted and are assigned to an appropriate border number; refer to Appendix and Figure A1 for details on the algorithmic procedures. As a note on computation, it takes many hours for the server-like machine to have the process done: France and UK each about 5 hours, and Germany 63 hours. Eventually those dangling arcs are dropped out in the process of extracting the blocks.

The border number is a de facto topological distance of a block far from the outmost border (Note: the border is not necessarily a country border). Every block has a border number, showing how far it is from the outmost border. The higher the border number, the farther the block is from the border, or reversely the lower the border number, the closer the block is to the border.

We apply the above computing process to the three networks, and obtain large numbers of blocks as shown in Table 1. We can note that Germany has far more blocks than France and UK, almost four times as many. Each individual block is assigned a border number to show how far it is to the outmost border. Interestingly, the maximum border number for the farthest block from the border varies from one country to another. Both France and UK have very similar maximum border numbers at around 70-80, while Germany has the largest maximum border number of 125. We will illustrate that the blocks with the maximum border number tend to be in the deep center of the countries. By center, we mean topological centers rather than geometric ones. This point will be clearer later on when we visualize all the blocks according to their border numbers.

**3. Lognormal distribution of block sizes and head/tail division rule**
We conduct detailed analysis of the extracted blocks, and find that there are far more small blocks than large ones. For example, 90% of blocks are small ones, while 10% are large ones in the UK. This observation is applicable to the other two networks, although the percentages may slightly vary from one another (Table 2). The block sizes for the individual countries follow a lognormal distribution (Figure 2), rather than a power law distribution as claimed in a previous study by Lämmer et al. (2006) that focuses only on city blocks. This finding is very interesting. Even more interesting is the fact that the mean of block sizes can make a clear-cut distinction between small blocks and large blocks. For example, 0.4 square kilometers is the mean to distinguish the small blocks and the large blocks for the UK network. This mean varies slightly from one country to another (Table 2). Overall, there is a very high percentage of small blocks and a very low percentage of large blocks. Put more generally, *given a variable X, if its values x follow a heavy tailed distribution, then the mean (m) of the values can divide all the values into two parts: a high percentage in the tail, and a low percentage in the head.* For the sake of convenience, we call this regularity the head/tail division rule.

Table 2: The numbers of blocks in head and tail for the three networks

|  | France | Germany | UK |
|---|---|---|---|
| Number of blocks (all) | 569,739 | 2,095,388 | 586,809 |
| Mean value of block sizes (km$^2$) | 0.9 | 0.2 | 0.4 |
| Number of blocks (< mean) | 534,800 | 1,802,924 | 526,302 |
| Tail (< mean) % | 94% | 86% | 90% |
| Head (> mean) % | 6% | 14% | 10% |

We apply the head/tail division rule into the three networks, and find some very interesting results related to the 80/20 principle or the principle of least effort (Zipf 1949). As seen from Table 2, about 10% of blocks are larger blocks constituting rural areas, and 90% of blocks are smaller blocks forming urban areas. On the other hand, the 10% of blocks enclose about 90% of the land, and the 90% of the blocks enclose only 10% of the land, which is located in cities. Reversely, the 10% of the land owns 90% of the blocks (which are in cities), and the 90% of the land owns only 10% of the blocks (which are in the countryside). This inbuilt imbalance between the minority (in the head) and the majority (in the tail) is exactly what the 80/20 principle or the principle of least effort illustrates. A further investigation on the areas of the blocks indeed illustrates the above facts as shown in Table 3, where all the blocks are divided into the head and the tail in terms of the areas of the blocks rather than the number of blocks. The head/tail division rule constitutes a foundation for delineating cities which is the topic of the next section.

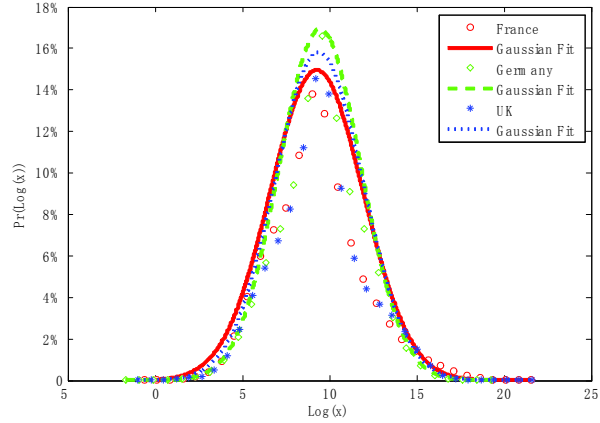

Figure 2: (Color online) Lognormal distribution of the block sizes for the three street networks

Table 3: The areas (km$^2$) of blocks in the head and the tail for the three networks

|  | France | Germany | UK |
| --- | --- | --- | --- |
| Area of blocks (all) | 528,706 | 353,706 | 209,062 |
| Mean of block sizes | 0.9 | 0.2 | 0.4 |
| Area of blocks (< mean) | 27,687 | 42,669 | 16,186 |
| Head (< mean) % | 5% | 12% | 8% |
| Tail (> mean) % | 95% | 88% | 92% |

**4. Delineating urban boundaries or defining natural cities based on head/tail division rule**

Because all blocks in one country exhibit a heavy tailed distribution, we can use the mean to divide all the blocks into smaller ones (smaller than the mean) and larger ones (larger than the mean). We then cluster the smaller blocks into individual groups. This clustering process goes like this. Starting from any smaller block whose neighboring blocks are also smaller ones, we design a program to traverse its adjacent blocks, and cluster those smaller blocks whose adjacent blocks are also smaller ones. This processing continues recursively until all the smaller ones are exhausted. We find that the sizes of the clustered groups demonstrate a heavy tailed distribution. Because of this, we then rely on the head/tail division rule to divide the groups into smaller ones (smaller than the mean) and larger ones (larger than the mean). The larger groups are de facto cities or natural cities as shown in Figure 3.

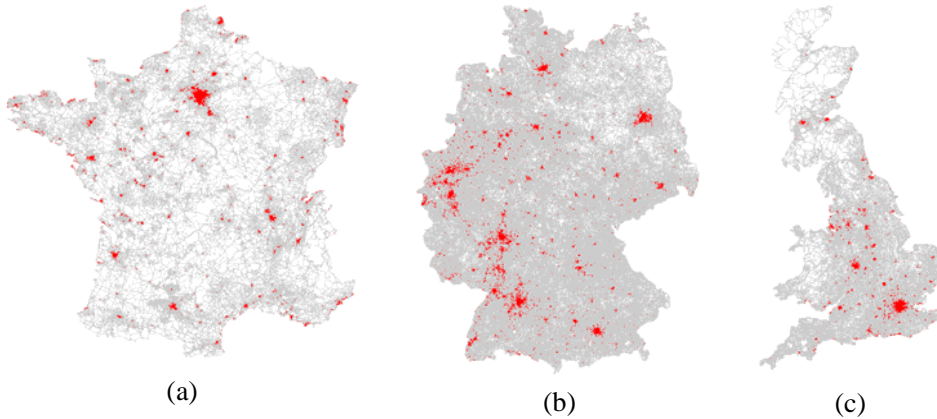

(a)          (b)          (c)

Figure 3: (Color online) All natural cities in red identified for the three networks of (a) France, (b) Germany, and (c) UK (Note: the gray background shows the extracted blocks)

The natural cities in this paper are defined as agglomerated smaller blocks (smaller than the mean) whose adjacent blocks are also smaller ones. In the above clustering process, we in fact consider the spatial autocorrelation effect, which helps to exclude smaller blocks whose adjacent blocks are larger ones. The spatial autocorrelation can be expressed equivalently by the first law of geography, which reads "*everything is related to everything else, but near things are more related than distant things*" (Tobler 1970). These smaller blocks, whose adjacent blocks are larger ones, are unlikely to be part of cities. Following this rule has one potential side effect, i.e., the smaller blocks that initially form a part of a city edge are mistakenly categorized into the countryside category. However, this side effect is very trivial. This is because the blocks that are mistakenly categorized into countryside are at the city edge, so they are relatively large according to spatial autocorrelation. So categorizing them as countryside is still a reasonable decision given their relatively large sizes.

The above process of delineating city boundaries is pretty straightforward. It is mainly the selection based on the head/tail division rule and a clustering process of smaller blocks by considering spatial autocorrelation. The entire process can be summarized as this: all blocks > smaller blocks > clustered groups > cities, where > denotes the processes of the selection or clustering. It is worth noting that only the head part of the clustered groups is considered to be cities, i.e., we filtered out smaller groups in the tail. These cities are what we call natural cities, as the boundaries are naturally or automatically obtained. The sizes of the natural cities follow power law distributions as shown in Figure 4. This power law detection is strictly based on the method suggested by Clauset et al. (2009), but only France and Germany pass the goodness of fit test. The reason why the UK did not pass the statistical test is a matter of research. However, we suspect that it may be due to the fact that it is an island country, and the urban grown is constrained by the coastline.

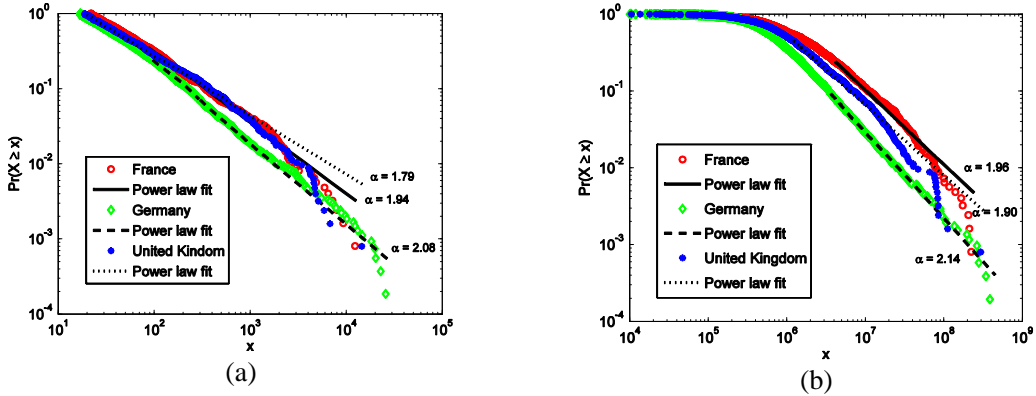

Figure 4: (Color online) Power law distribution of city sizes for the three networks in terms of (a) the number of city blocks, and (b) the area or physical extent (m$^2$)

### 5. Mapping the image of the country using the border number
In this section, we will examine how the defined border number in Section 2 can be used to map the image of the country, a notion with a similar meaning as in *The Image of the City* (Lynch 1960). That is how those distinguished cities (being landmarks) shape the mental image of the country by filtering out the vast majority of redundant or trivial human settlements in the heavy or fat tail. As we have learned, all those blocks on the border has border number 1, those blocks adjacent to the border number 1 have border number 2, so on and so forth. Every block has a unique border number, the more central a block, the higher the border number. The blocks with the highest numbers are within the deep centers of the megacities in the countries. Using a spectral color legend, we map the blocks according to their border numbers (Figure 5). The patterns shown in Figure 5 reflect the true structure of the countries, e.g., the red spots or patches are the true centers of the countries. The sizes of the spots or patches vary from one country to another. This reflects another interesting structure which deserves some further elaboration. We learn from the mapping that the red spots or patches represent the largest cities (in terms of both the number of blocks and physical extent) in the countries. Due to the use of Jenks' natural breaks (Jenks 1967), which minimize intra-classes variation and maximize inter-classes one, it implies that in France there is no other cities in the same group as Paris – the largest city in the country. This is indeed true that Paris has 62,242 city blocks, more than five times bigger than the second largest city. In the case of Germany, this is another extreme that all biggest cities are very similar in size around 20 to 30 thousand blocks. This can also be seen from Figure 3.

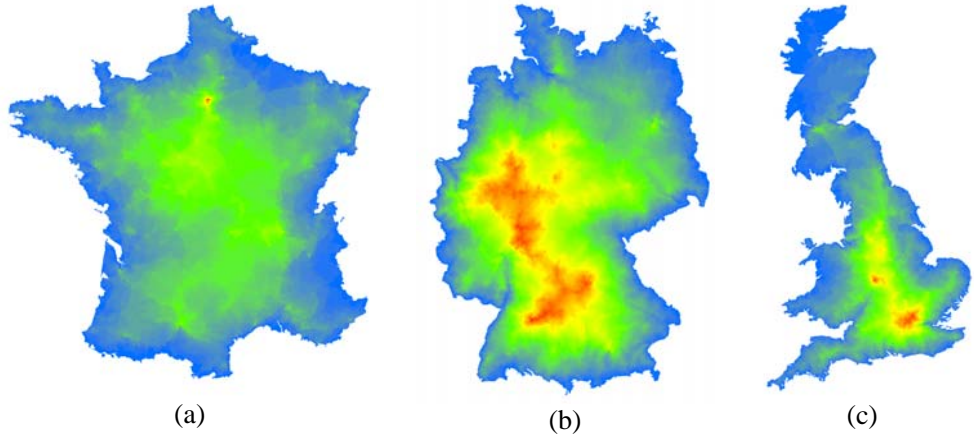

(a) (b) (c)

Figure 5: (Color online) Mapping the border number using a spectral color legend
(Note: the higher the border number the warmer the color; red and blue represent respectively the highest and lowest border numbers)

The patterns shown in Figure 5 are illustrated from a topological perspective, which is very different from a geometric one. For example, given any country border or shape, we can partition the shape into equal sized rectangular cells (at a very fine scale, e.g., 1000 x 1000), and then compute the border number for the individual cells. Eventually, we obtain the patterns shown in Figure 6. As we can see, the centers of the countries are geometric or gravity centers that are equal distances to the corresponding edges of the borders. Essentially the country forms or shapes are viewed symmetrically. This is a distorted image of the countries, since the geometric centers are not true centers that the human minds perceive. This geometric view is the fundamental idea behind the concept of medial axis (Blum 1967), which has found a variety of applications in the real world in describing the shape of virtually all kinds of objects from the infinitely large to the infinitely small including biological entities (Leymarie and Kimia 2008). While medial axis is powerful enough in capturing a symmetric structure of a shape, it presents a distorted image of a shape as seen from Figure 6. This distortion is particularly true for France, since the true center Paris is far from the geometric or gravity center. We should stress that the point here is not so much about how Figure 5 captures the mental image of the country for different people, BUT rather than how the topological perspective (the border number and block perspective) is superior to the geometric view as shown in Figure 6, which gives a distorted image. This is the message we want to convey about the image of the country.

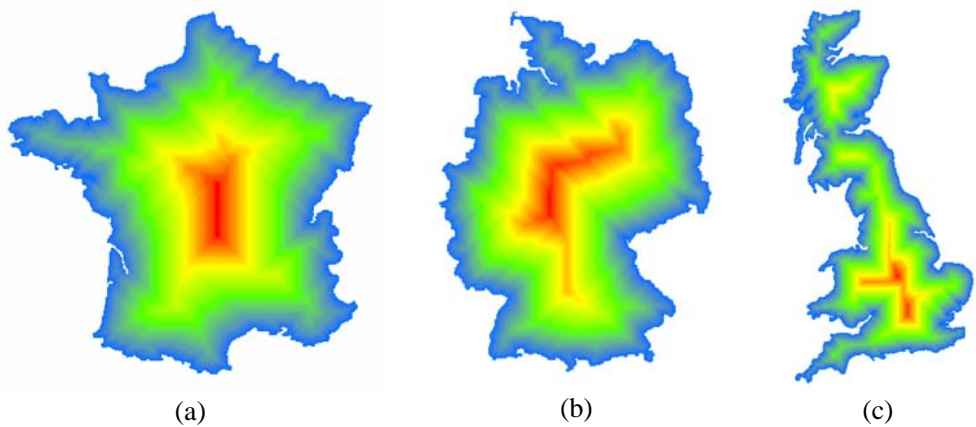

(a) (b) (c)

Figure 6: (Color online) Distorted images of the country based on the geometric distance far from the outmost boundaries

The above scaling analysis and insights into the structure can be extended to individual cities to illustrate their internal structure and patterns. Taking London for example, those blocks with the highest border numbers reflect a true center of the city (Figure 7a). We found that the city block sizes of London follow a lognormal distribution as well, one of the heavy tailed distributions. To some extent, it is consistent with an early study that claims a power law distribution of city block sizes (Lämmer et al. 2006), since both lognormal and power law are heavy tailed distributions. The fact that the block sizes follow a lognormal distribution gives us a possibility to distinguish the blocks above the mean and below the mean. Those blocks above the mean constitute individual city core or cores, which match pretty well to the city center (Figure 7b).

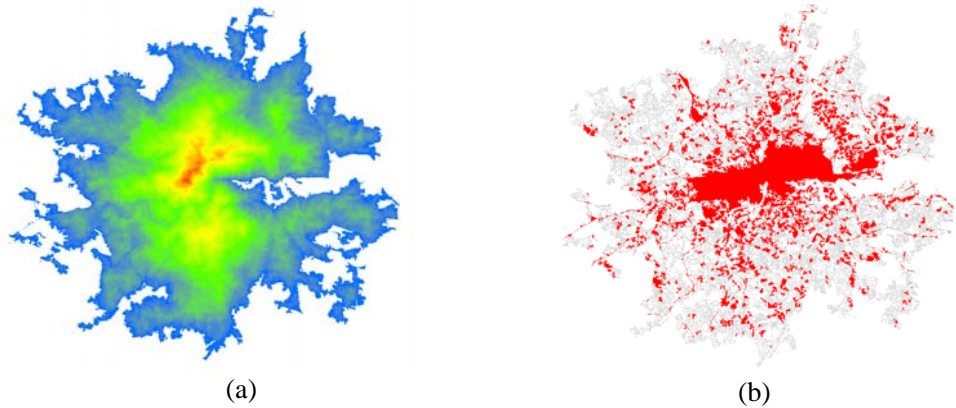

(a)  (b)

Figure 7: (Color online) Scaling patterns illustrated with the case of London (a) city border number, and (b) city cores (Note: the border is relative to the London city rather than the UK country border)

**6. Discussions on the study**
This study provides an example of uncovering new knowledge from massive geographic information using ever increasing computing power of personal computers. The increasingly available volunteered geographic information such as the OSM data, GPS traces, and geotagged photos provides an unprecedented data source for the kind of scaling analysis without any particular constraints (Crandall et al. 2009). Nowadays researchers can take the street networks of virtually the entire world for computing and analysis. The computation at the massive data scale can lead to some in-depth insights into structure or patterns of large geographic space. For example, it is a surprise to us that the head/tail division rule can be used to delineate city boundaries in such a simple manner. The ever increasing computing power of personal computers makes the data-intensive computing involved in this kind of study possible. As a by-product of the study, we generate two valuable datasets that can be of use for various urban studies: one about blocks and another about natural cities for the three counties. We will release soon the data up to 2 GB for research uses.

The head/tail division rule we derived from the study has some implications to other natural and socio-economic phenomena that exhibit a power law distribution or a heavy tail distribution in general. For example, the distinction between rich and poor may be clearly defined by the mean of all individuals' wealth in a country or society. This of course warrants a further study for verification. In a previous study (Jiang 2007), we divided streets around the mean connectivity (which is 4) into two categories: well-connected and less connected. In addition, those extremely well connected or extremely few tend to shape part of mental maps about geographic space or cities. Those most connected streets in a city tend to stay in the human mind, just as most people tend to remember those richest people in the world.

We believe that the perspective of city and field blocks has a certain implication to understanding morphology of organism as well. This perspective of city and field blocks is in fact a topological perspective rather than a geometric one. This can be seen from the difference of pattern or structure illustrated by Figures 5 and 6: the former being the topological perspective, while the latter being the geometric one. Given a biological entity like a human body, the medial axis can indeed help to derive bone structure or skeleton which is symmetric in essence, just like the one illustrated in Figure 6. However, it is hard or impossible for medial axis to derive true centers of human bodies: the brains and the hearts. Consequently, we believe that the idea of the border number can be extended to some complex organisms. We know that cells are the basic constituents of biological organisms. For example, human beings have about 100 trillion or $10^{14}$ cells, and the cell sizes range from 4 - 135 µm. We believe based on the high ratio of the biggest cell to the smallest cell 135/4 that the human body's cell sizes follow a heavy tailed distribution. Also the highly complex organisms like the brains and the hearts tend to have smaller cells. This is much like the smaller city blocks in a capital city or other megacities in a country. Thus mapping the border number for all the human body cells that make up the solid tissues including muscle and skeletal cells would illustrate true centers of the human body, i.e., two red spots representing the heart and the mind as one can imagine following the patterns shown in Figures 5 and 7. Note that the cells we discussed and compared here could be unknown units, which are equivalent to the blocks in geographic space. This speculation is not yet supported by scientific references. We believe also that for the cells or the unknown units of the brains and hearts they have similar heavy tailed distributions of sizes as the cells in the human bodies.

Following the analogue between geographic space and complex organisms, we tend to believe that geographic spaces or cities in particular are self-organized. There has been an increasing awareness that a city needs to be better understood as a biological entity in terms of its form and how it functions. For example, theoretical physicist Geoffrey West and his co-workers (1999) have studied a set of scaling laws originated in biology and found that the laws hold true remarkably precise for cities. The scaling laws state a nonlinear relationship, e.g., the bigger the city, the less infrastructure per capital. This economy of scale indicates that various infrastructures such as utilities networks, street networks and buildings needed to sustain a big city are far less than the amount one would expect. This is much along the line of research on allometry - the study on the growth of part of an organism in relation to that of the entire organism. Some studies on how cities can be compared to organisms have been carried out (Samaniego and Moses 2008, Batty et al. 2008, Steadman 2006). We can foresee more studies coming up in the near future due to the availability of massive geographic information.

## 7. Conclusion

Geographic space is essentially very large and very complex. There is no average place that can represent any other places. The scaling of geographic space can characterize this inherent "spatial heterogeneity" by a heavy tailed distribution. In this paper, we examined the scaling of geographic space from the perspective of city and field blocks. We found that block sizes at both country and city levels exhibit a lognormal distribution, and that the mean size can divide all the blocks into a low percentage at the head and a high percentage in the tail. This head/tail division rule has been used to differentiate urban and rural areas. Thus we developed a simple solution to delineating urban boundaries from large street networks. The perspective of blocks is unique in the sense that it can capture underlying structure and patterns of geographic space. In this regard, the defined border number is particularly of use in detecting the centers of a country or a city.

This study further adds some implications to understanding the morphology of organisms. The city and field blocks can be compared to the cells of complex organisms. We believe that this kind of scaling analysis of geographic space can be applied to complex organisms and we consequently conjecture that a similar scaling structure is appeared in complex organisms like human bodies or human brains. This would reinforce our belief that cities, or geographic space in general, can be compared to a biological entity in terms of their structure and their self-organized nature in their evolution. Our future work will concentrate on the further verification of the findings and applications of the head/tail division rule.

**Acknowledgement**
We thank the OSM community for providing the impressive data of street networks. We also want to thank Tao Jia for his assistance in creating Figure 2, and Marc Barthelemy for his insightful comments on the head/tail division rule. All source codes and produced data are put online for free access at http://fromto.hig.se/~bjg/scalingdata/.

**Appendix: Algorithmic functions and flow chart for computing individual blocks and the border numbers**

This appendix is to supplement the description of the data processing introduced in Section 2. The algorithmic functions and flow chart provide a very detailed description about the computational processes. Interested readers are encouraged to contact us for access of the source codes.

```
Input: a set of arcs with calculated deflection angles
Output: block set
Function BlockDetection ()
    Select the rightmost (leftmost, uppermost or downmost) one in arcs as the start arc
    Let maximum block = MaximumBlock (start arc)
    Let border number of maximum block = 0
    Let new block list = null
    For each arc in maximum block
        Let current block = BlockTracking (current arc, the traversal strategy)
        Let border number of current block = 1
        Add current block to new block list
    While (the number of blocks in new block list > 0)
        Let base block = the first block in new block list
        Add base block to block set
        For each arc in base block
            Let current block = BlockTracking (current arc, the opposite traversal strategy)
            Let border number of current block = border number of base block + 1
            Add current bock to new block list
        Remove the first block in new block list
    Return block set

Function MaximumBlock (start arc)
    Let left block = BlockTracking (start arc, left)
    Let right block = BlockTracking (start arc, right)
    Let maximum block = null
    If (left block contains more of the arcs)
        Let maximum block = left block
        Let traversal strategy of maximum block = left
    If (right block contains more of the arcs)
        Let maximum block = right block
        Let traversal strategy of maximum block = right
    Return maximum block

Function BlockTracking (start arc, traversal strategy)
    Let next arc = null
    Let current arc = start arc
     While (next arc != start arc)
        If (traversal strategy is right)
           Select rightmost connected one in arcs connected with current arc as next arc
        Else
           Select leftmost connected one in arcs connected with current arc as next arc
        Add current arc to current block
        Let current arc = next arc
    Return current block
```

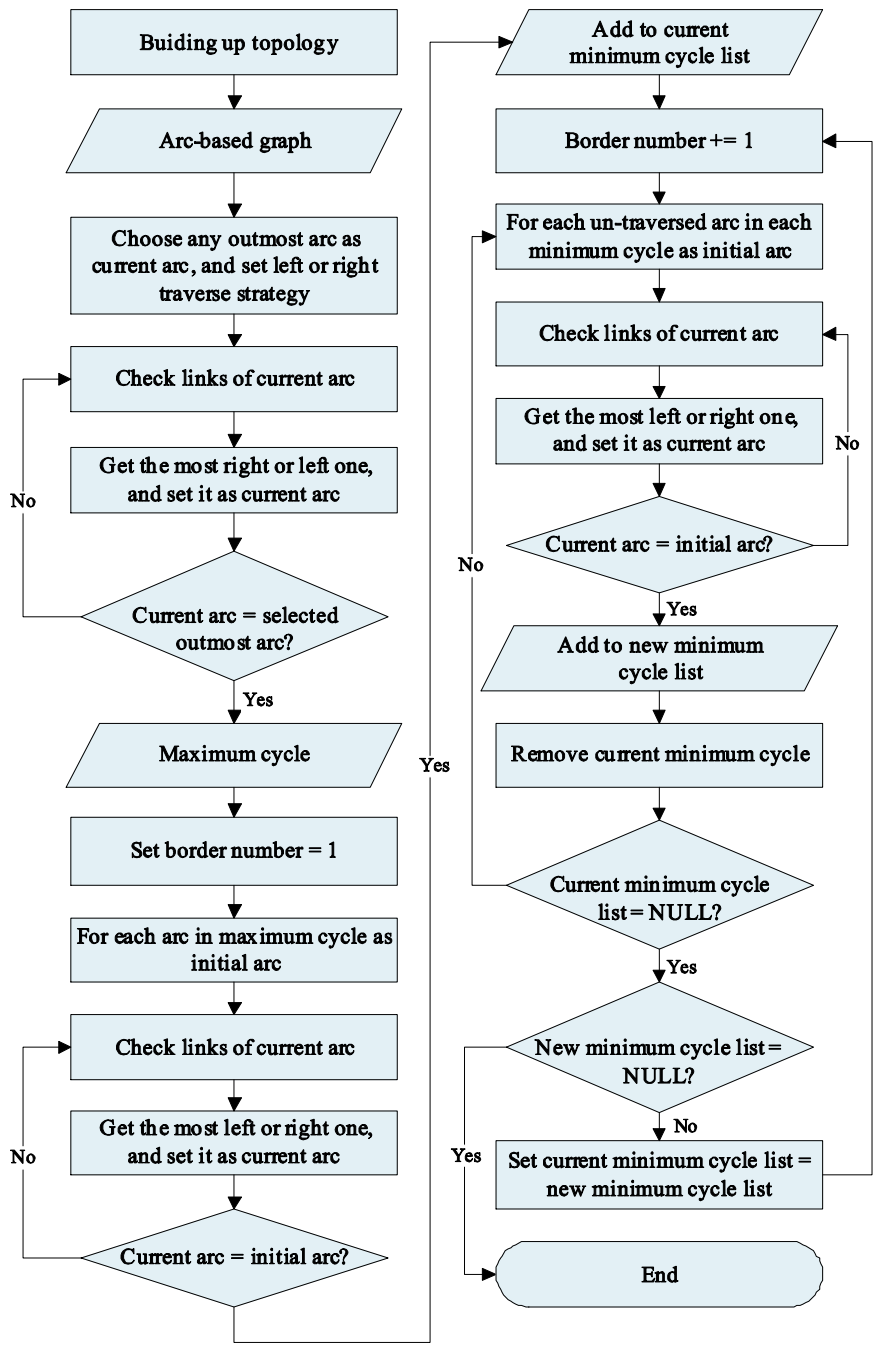

Figure A1: Flow chart of the data processing for computing individual blocks and their border numbers